\documentclass[12pt,fleqn]{article}
\usepackage{graphicx}

\begin{document}

\newcommand{\rf}[1]{(\ref{#1})}
\newcommand{\rff}[2]{(\ref{#1}\ref{#2})}

\newcommand{\ba}{\begin{array}}
\newcommand{\ea}{\end{array}}

\newcommand{\be}{\begin{equation}}
\newcommand{\ee}{\end{equation}}

\newcommand{\const}{{\rm const}}
\newcommand{\ep}{\varepsilon}
\newcommand{\Cl}{{\cal C}}

\newcommand{\e}{{\bf e}}

\newcommand{\m}{\left( \ba{c}}
\newcommand{\ema}{\ea \right)}
\newcommand{\mm}{\left( \ba{cc}}
\newcommand{\miv}{\left( \ba{cccc}}

\newcommand{\scal}[2]{\mbox{$\langle #1 \! \mid #2 \rangle $}} 
\newcommand{\ods}{\par \vspace{0.5cm} \par}

\newtheorem{prop}{Proposition}
\newtheorem{Th}{Theorem}  
\newtheorem{lem}{Lemma}
\newtheorem{rem}{Remark}
\newtheorem{cor}{Corollary}
\newtheorem{Def}{Definition}
\newtheorem{open}{Open problem}
\newtheorem{ex}{Example}
\newtheorem{exer}{Exercise}

\title{\bf On simulations of the classical harmonic oscillator 
equation by difference equations}

\author{
 {\bf Jan L.\ Cie\'sli\'nski}\thanks{\footnotesize 
 e-mail: \tt janek\,@\,alpha.uwb.edu.pl}
\\ {\footnotesize Uniwersytet w Bia\l ymstoku,  
Instytut Fizyki Teoretycznej}
\\ {\footnotesize ul.\ Lipowa 41, 15-424  
Bia\l ystok, Poland}
\\ {\bf Bogus\l aw Ratkiewicz}\thanks{\footnotesize
e-mail: \tt bograt\,@\,poczta.onet.pl}
\\ {\footnotesize I Liceum Og\'olnokszta\l c\c ace, 16-300 August\'ow,
Osiedle \'Sr\'odmie\'scie 31, Poland}
\\ {\footnotesize Doctoral Studies, Wydzia\l \ Fizyki, Uniwersytet Adama Mickiewicza, Pozna\'n, Poland}
}

\date{}

\maketitle

\begin{abstract} 
We show that any second order linear ordinary diffrential equation 
with constant coefficients (including the damped and undumped harmonic oscillator equation)
admits an exact discretization,
i.e., there exists a difference equation whose solutions exactly coincide 
with 
solutions of the corresponding differential equation evaluated at 
a discrete sequence of points (a lattice). Such exact discretization 
is  found for an arbitrary lattice spacing.
\end{abstract}

\vspace{0.3cm}

\section{Introduction}
The motivation for writing this paper is an observation that 
small and apparently not very important changes in the 
discretization of a differential equation lead to 
difference equations with completely different properties.
By the discretization we mean a simulation of the 
differential equation by a difference equation \cite{Hi}.

In this paper we consider  
 the damped harmonic oscillator equation
\be \label{damped-os}
  \ddot x + 2 \gamma \dot x + \omega_0^2 x = 0 \ .
\ee
where $x= x(t)$ and the dot means the $t$-derivative. 
This is a linear equation and its general solution is well known.
Therefore discretization 
procedures are not so important (but sometimes are applied, see \cite{Souza}). 
However, this example allows us to show and illustrate some more 
general ideas.

The most natural discretization, known as the Euler method (Appendix~\ref{app-numer}, 
compare \cite{Hi,Po}) consists in replacing $x$ by $x_n$, $\dot x$ by 
the difference ratio $(x_{n+1}-x_n)/\ep$, $\ddot x$ by the difference ratio
of difference ratios, i.e.,
\be  \label{ddot}
  \ddot x \quad \rightarrow  \quad \frac{1}{\ep} \left( \frac{x_{n+2} - x_{n+1}}{\ep} -
 \frac{x_{n+1}-x_n}{\ep} \right) =  
\frac{ x_{n+2} - 2 x_{n+1} + x_{n} }{\ep^2} \ ,
\ee
and so on. This possibility is not unique. We can 
replace, for instance, $x$ by $x_{n+1}$, or $\dot x$ by 
$(x_{n}-x_{n-1})/\ep$, or $\ddot x$ by 
$(x_{n+1} - 2 x_{n} + x_{n-1})/\ep^2$. Actually the last formula, due to its symmetry, seems to
be more natural than \rf{ddot} (and it works better indeed, see Section~\ref{three}).

In any case we demand that the continuum limit, i.e.,
\be  \label{continuum}
x_n = x(t_n) \ , \qquad t_n = \ep n \ , \qquad \ep \rightarrow 0 \ ,
\ee
applied to any discretization of a differential equation yields this  
differential equation.
The continuum limit consists in replacing $x_n$ by $x(t_n) = x(t)$ and the
neighbouring values are computed from the Taylor expansion of the function 
$x(t)$ at $t=t_n$
\[
 x_{n + k} = x (t_n + k \ep) =  x (t_n) + {\dot x} (t_n) k \ep 
+ \frac{1}{2} {\ddot x} (t_n) k^2 \ep^2 + \ldots \ 
\]
Substituting these expansions into the difference equation 
and leaving only the leading term we should obtain the considered  
differential equation.

In this paper we compare various discretizations of the damped 
(and undamped) harmonic oscillator equation.
The main result of the present paper consists in finding the exact discretization
of the damped harmonic oscillator equation \rf{damped-os}. 
By exact discretization we mean that  $x_n = x (t_n)$  holds for any $\ep$ and
not only in the limit \rf{continuum}.

\section{Simplest discretizations of the harmonic oscillator}
\label{three}

\begin{figure}[p]  
\caption{\small Simplest discretizations of the harmonic oscillator equation for small $t$ 
and $\ep =0.02$. 
Black points: solution of Eq.~\rf{dis-Euler-os}, white points: Eq.~\rf{dis-sym-os}, 
grey points: Eq.~\rf{dis-forw-os}, thin line: exact continuous solution.  }  
\label{simple-small} \par
\includegraphics[width=\textwidth]{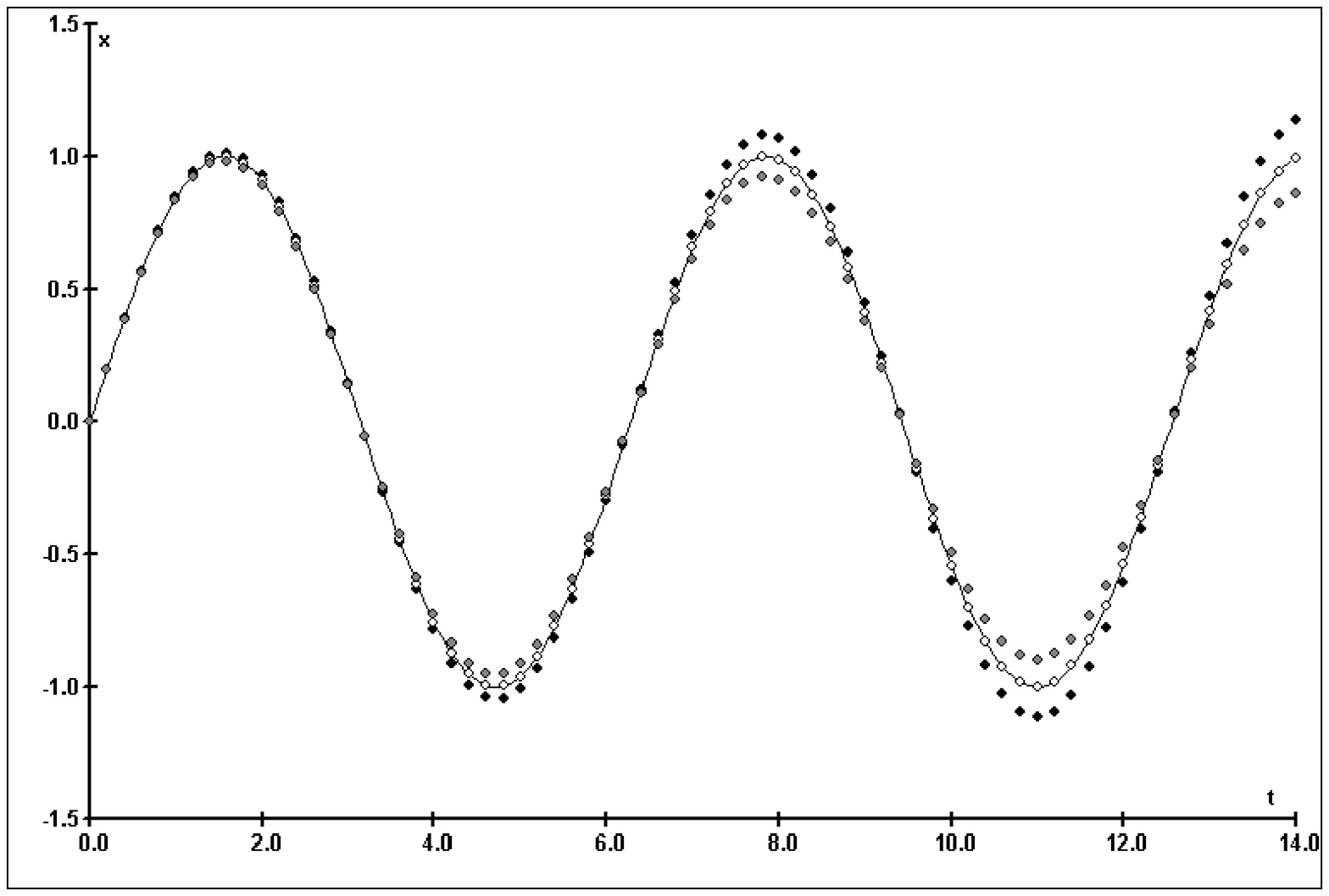} \par

\end{figure}

\begin{figure}[p]  
\caption{\small Simplest discretizations of the harmonic oscillator equation for large $t$ 
and $\ep =0.02$.
Black points: solution of Eq.~\rf{dis-Euler-os}, white points: Eq.~\rf{dis-sym-os}, 
grey points: Eq.~\rf{dis-forw-os}, thin line: exact continuous solution. }
 \label{simple-large} \par
 \includegraphics[width=\textwidth]{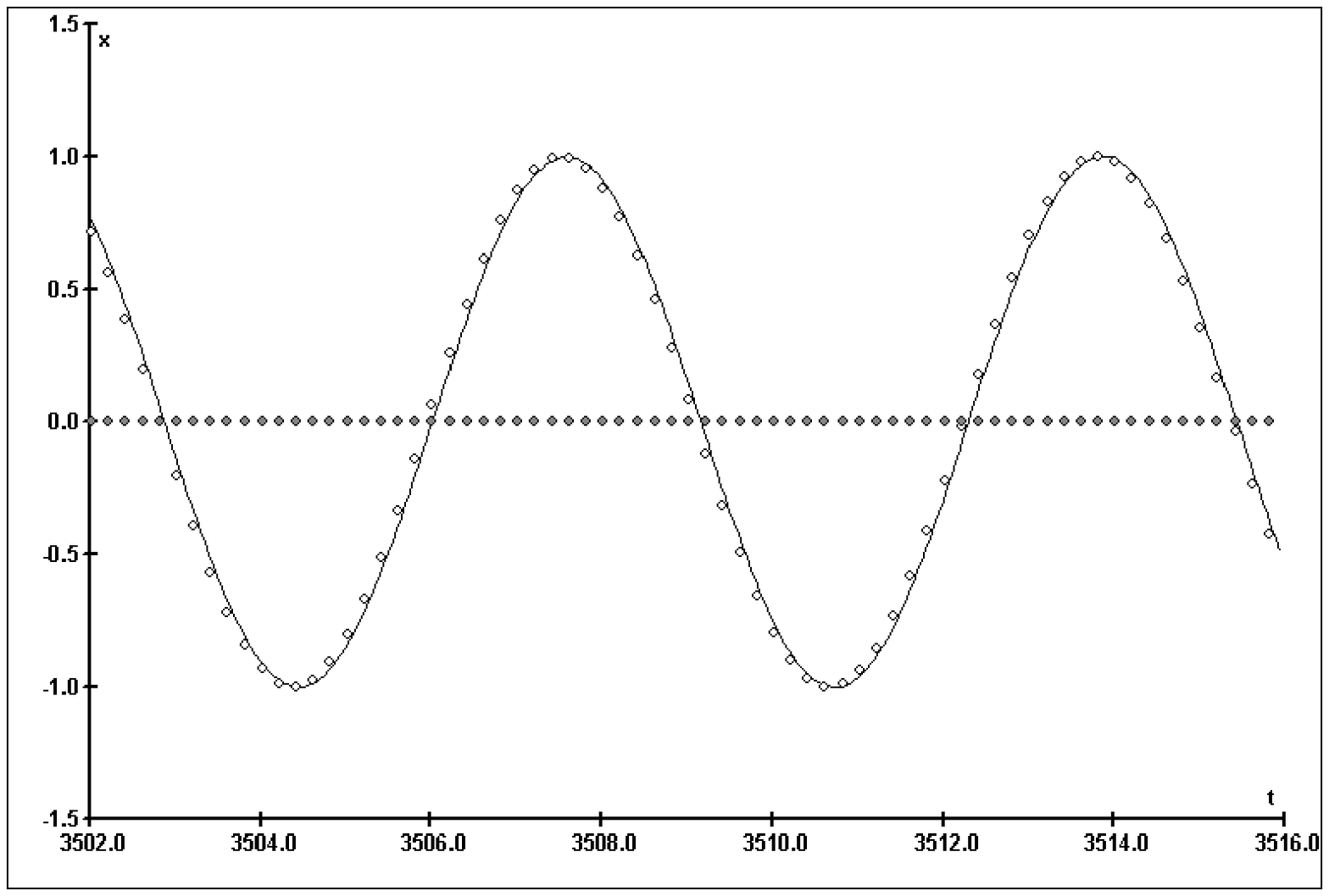} 
\end{figure}

Let us consider the following three discrete equations 
\be  \label{dis-Euler-os}
  \frac{ x_{n+1} - 2 x_{n} + x_{n-1} }{\ep^2} + x_{n-1} = 0 \ ,
\ee \par
\be  \label{dis-sym-os}
  \frac{ x_{n+1} - 2 x_{n} + x_{n-1} }{\ep^2} + x_{n} = 0 \ ,
\ee \par
\be  \label{dis-forw-os}
  \frac{ x_{n+1} - 2 x_{n} + x_{n-1} }{\ep^2} + x_{n+1} = 0 \ ,
\ee 
where $\ep$ is a constant. The continuum limit \rf{continuum} 
 yields, in any of these cases, the harmonic oscillator 
equation 
\be  \label{oscyl}
       \ddot x + x = 0 \ . 
\ee
To fix our attention, in this paper we consider  only solutions corresponding to 
the initial conditions $x(0)=0$, ${\dot x} (0) = 1$. The initial data for the discretizations
are chosen in the simplest form: we assume that $x_0$ and $x_1$ belong to the graph of the 
exact continuous solution. 

For small $t_n$ and small $\ep$ the discrete solutions of any of 
these equations 
approximate the corresponding continuous solution quite well
(see Fig.~\ref{simple-small}). However, 
the global behaviours of the solutions (still for small $\ep$!) are 
different (see Fig.~\ref{simple-large}).  The solution of the equation \rf{dis-forw-os}
vanishes at $t \rightarrow \infty$ while the solution of \rf{dis-Euler-os} 
oscillates with rapidly incrising amplitude (all black points  
are outside the range 
of Fig.~\ref{simple-large}).
Qualitatively, only the discretization \rf{dis-sym-os} resembles the 
continuous case. However, for very large $t$ 
 the discrete solution becomes increasingly different from the exact continuous solution 
even in the case \rf{dis-sym-os} 
(compare Fig.~\ref{simple-large} and Fig.~\ref{good-smep}).  

The natural question arises {\it how to find a discretization which is the 
best as far as global properties of solutions are concerned}.

In this paper we will show how to find the ``exact'' discretization of 
the damped harmonic oscillator equation. 
In particular, we will present the discretization of \rf{oscyl} which is 
better than \rf{dis-sym-os} and, in fact, seems to be the best possible.
We begin, however, with a very simple example which illustrates 
the general idea of this paper quite well.

\section{The exact discretization of the exponential growth equation}

 \label{firstorder}

We consider the discretization of the equation
$   \dot x = x $.
Its general solution reads
\be  \label{consol1}
x (t) = x(0) e^t \ .
\ee
The simplest discretization is given by
\be  \label{dis1}
\frac{x_{n+1} - x_n}{\ep} = x_n \ .
\ee
This discrete equation can be solved immediately. Actually this is just
the geometric sequence  $x_{n+1} = (1 + \ep) x_n$.
 Therefore
\be \label{dissol1}
      x_n = (1+\ep)^n x_0 \ .
\ee
To compare with the continuous case we write 
$(1+\ep)^n$ in the form
\be
    (1+\ep)^n = \exp ( n \ln (1+\ep) ) 
 = \exp (\kappa t_n) \ ,
\ee
where $t_n : = \ep n$ and  $\kappa := \ep^{-1} \ln (1 + \ep)$.
Thus the solution \rf{dissol1} can be rewritten as 
\be  \label{dissol11}
    x_n = x_0 e^{\kappa t_n} \ .
\ee
Therefore we see that for $\kappa \neq 1$  
the continuous solution \rf{consol1}, evaluated at
$t_n$, i.e. 
\be  \label{exact1}
x(t_n) = x(0)  e^{t_n} \ , 
\ee
differs from the corresponding discrete 
solution \rf{dissol11}.
One can easily see that $0  < \kappa < 1$. Only in the limit
$\ep \rightarrow 0$ we have $\kappa \rightarrow 1$.

Although the qualitative behaviour of the ``naive'' simulation \rf{dis1}
is in good agreement with the smooth solutions (exponential growth in both
cases) but quantitatively the discrepancy is very large for 
$t\rightarrow \infty$ because the exponents are different.

The discretization \rf{dis1} can be easily improved. Indeed, 
replacing in the formula \rf{dissol1} $1+\ep$ by $e^\ep$ 
we obtain that \rf{dissol1} coincides with the exact solution \rf{exact1}.
This ``exact discretization'' is given by:
\be \label{dis1-exact}
 \frac{ x_{n+1} - x_{n} }{ e^{\ep} - 1} = x_n \ ,
\ee
or, simply, $x_{n+1} = e^{\ep} x_n$.
Note that $e^\ep \approx 1 + \ep$ (for $\ep \approx 0$)
and this approximation yields the equation \rf{dis1}.

\section{Discretizations of the harmonic oscillator: exact solutions}

The general solution of the harmonic oscillator equation
 \rf{oscyl} is well known
\be  \label{contsol}
   x(t) = x(0) \cos t +  {\dot x}(0) \sin t \ .
\ee
In Section~\ref{three} we compared graphically this solution with the simplest 
discrete simulations: \rf{dis-Euler-os}, \rf{dis-sym-os}, \rf{dis-forw-os}. 
Now we are going to present exact solutions of these discrete equations.

Because the discrete case is usually less known than the continuous one
we recall shortly that the simplest approach consists in searching 
solutions in the form $x_n = \Lambda^n$ (this is an analogue of the
ansatz $x (t) = \exp (\lambda t)$ made in the continuous case, for 
more details see Appendix~\ref{app-lineq}). As a result we get 
the characteristic equation for $\Lambda$. 

We illustrate this approach on the example of the equation \rf{dis-Euler-os}
resulting from the Euler method.
Substituting $x_n = \Lambda^n$ we get the following characteristic equation 
\be
    \Lambda^2 - 2 \Lambda + (1 + \ep^2) = 0 \ ,
\ee
with solutions $\Lambda_1 = 1 + i\ep$, $\Lambda_2 = 1 - i\ep$. The general 
solution of \rf{dis-Euler-os} reads 
\[
x_n = c_1 \Lambda_1^n + c_2 \Lambda_2^n \ ,
\]
and, expressing $c_1, c_2$ by the initial conditions $x_0, x_1$, we have
\[
x_n = x_1 \frac{(1+i\ep)^n - (1-i\ep)^n}{2i\ep} + x_0 
\frac{(1+ i \ep) (1-i\ep)^n - (1 - i \ep) (1+i\ep)^n }{2i\ep}  .
\]
Denoting
\be
      1 + i\ep = \rho e^{i\alpha} \ , 
\ee
where $\rho = \sqrt{1+\ep^2}$ and $\alpha = \arctan \ep$,
we obtain after elementary calculations
\be  \label{dissol0}
   x_n = \rho^n  \left( x_0 \cos(n \alpha) + \frac{x_1 - x_0}{\ep} 
\sin (n \alpha) \right) \ .
\ee
It is convenient to denote $\rho = e^{\kappa \ep}$ and
\be  \label{tkw}
     t_n = n\ep \ , \qquad 
  \kappa := \frac{1}{2\ep} \ln (1+\ep^2) \ , \qquad
 \omega := \frac{\arctan \ep}{\ep} \ ,
\ee
and then
\be  \label{dissol2}
    x_n = e^{\kappa t_n} ( x_0 \cos \omega t_n + \frac{x_1 - x_0}{\ep}
  \sin \omega t_n ) \ .
\ee
One can check  that $\kappa > 0$ and $\omega < 1$ for any $\ep > 0$.
For $\ep \rightarrow 0$ we have $\kappa \rightarrow 0$, 
$\omega \rightarrow 1$. Therefore the discrete solution \rf{dissol2}
is characterized by the exponential growth of the envelope amplitude and
a smaller frequency of oscillations than the corresponding 
continuous solution \rf{contsol}.

A similar situation is in the case \rf{dis-forw-os}, with only one (but very
important) difference: instead of the growth we have the exponential decay.
The formulas \rf{tkw} and \rf{dissol2} need only one correction to be valid 
in this case. Namely, $\kappa$ has to be changed to $ - \kappa$.

The third case, \rf{dis-sym-os}, is characterized by $\rho = 1$, and, therefore, 
the amplitude of the oscillations is constant (this case will be discussed 
below in more detail). 

These results  are in perfect agreement with the behaviour of
the solutions of discrete equations illustrated at Fig.~\ref{simple-small} 
 and Fig.~\ref{simple-large}.

\begin{figure}[p]  
\caption{\small Good discretizations of the harmonic oscillator equation for large $t$ 
and $\ep =0.02$.
Black points: exact discretization \rf{dis-os-ex}, white points: Eq.~\rf{dis-sym-os}, 
grey points: Runge-Kutta scheme \rf{GLRK-harm}, thin line: exact continuous solution.}

\label{good-smep}
 \includegraphics[width=\textwidth]{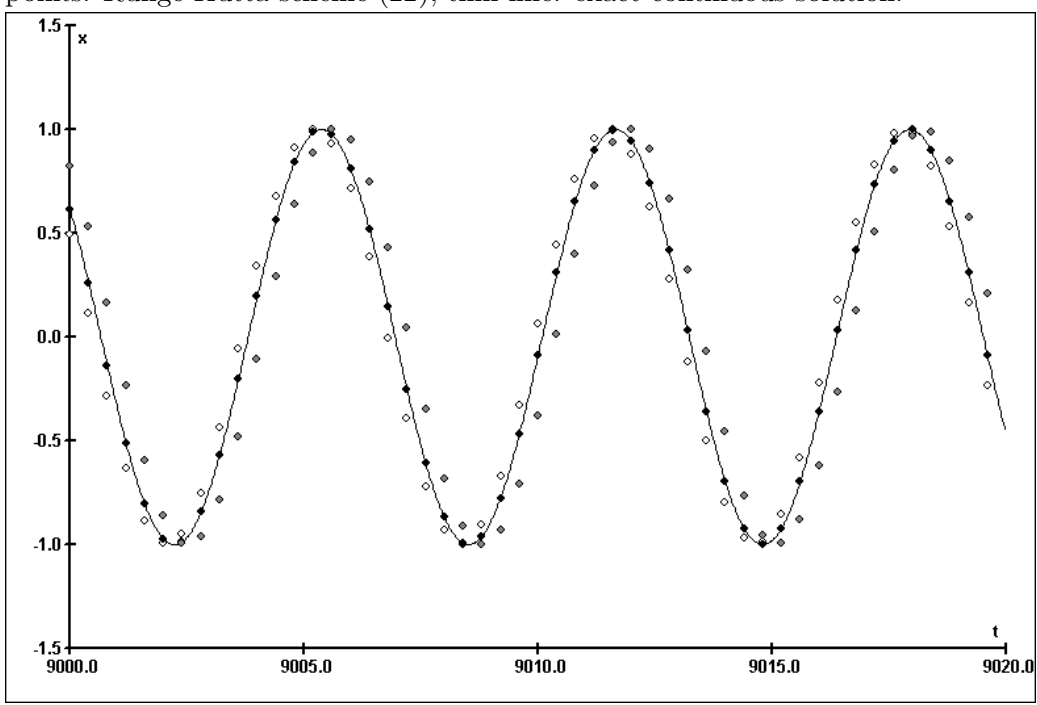} 

\end{figure}

\begin{figure}[p] 
\caption{\small Good discretizations of the harmonic oscillator equation for small $t$ 
and $\ep =0.4$.
Black points: exact discretization \rf{dis-os-ex}, white points: Eq.~\rf{dis-sym-os}, 
grey points: Runge-Kutta scheme \rf{GLRK-harm}, thin line: exact continuous solution. }

 \label{good-smt}
 \includegraphics[width=\textwidth]{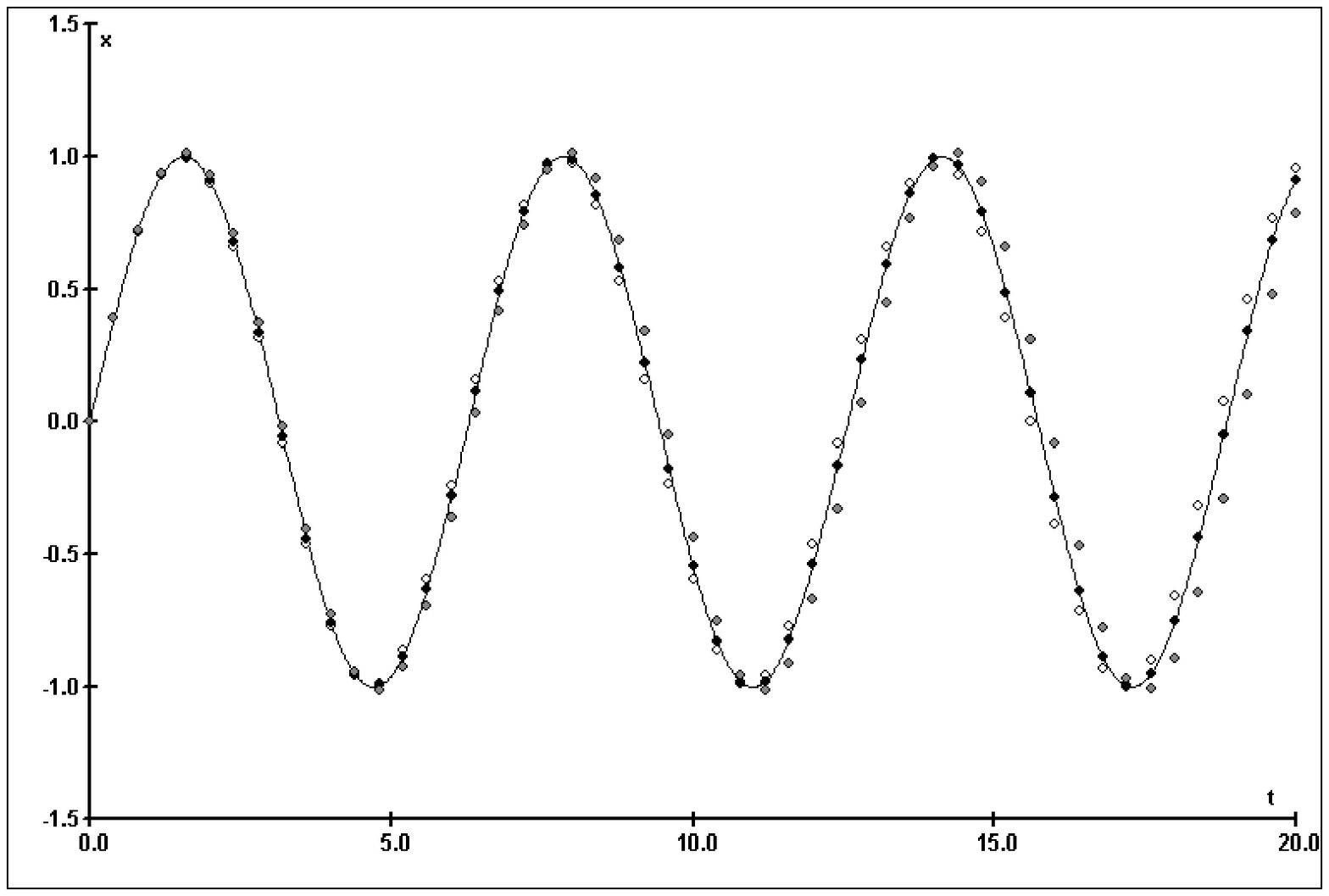} 

\end{figure}

Let us consider the following family of discrete equations
(parameterized by real parameters $p,q$):
\be  \label{family}
  \frac{ x_{n+1} - 2 x_{n} + x_{n-1} }{\ep^2} + p x_{n-1} + (1 - p - q)  x_n 
+ q x_{n+1} = 0 \ ,
\ee
The continuum limit \rf{continuum} applied to \rf{family}  
yields the harmonic oscillator \rf{oscyl}  for any $p, q$. The family \rf{family} contains all three  examples of Section~\ref{three} and (for $p=q=1/4$) the equation resulting from the Gauss-Legendre-Runge-Kutta method (see Appendix~\ref{app-numer}):
\be  \label{GLRK-harm}
x_{n+1} - 2 \left( \frac{ 4 - \ep^2}{4 + \ep^2} \right) x_n + x_{n-1} = 0 \ . 
\ee 
Substituting $x_n = \Lambda^n$ into \rf{family} we get the following characteristic equation:
\be \label{quadr}
(1 + q \ep^2) \Lambda^2 - ( 2  +  (p+q -1) \ep^2 ) \Lambda + 
(1 + p \ep^2)  = 0 
\ee

We formulate the following problem: {\it  find a discrete equation in the 
family \rf{family} with the  global behaviour of solutions as much similar to the continuous case as possible}.

At least two conditions seem to be very natural in order to get a ``good'' discretization 
of the harmonic oscillator: oscillatory character 
and a constant amplitude of the solutions (i.e., $\rho = 1$, $\kappa=0$). 
These conditions can be easily expressed in terms of roots ($\Lambda_1$, 
$\Lambda_2$) of the quadratic equation \rf{quadr}. First, the roots should be 
imaginary (i.e., $\Delta < 0$), second, their modulus should be equal to 1
i.e., $\Lambda_1 = e^{i \alpha}$, $\Lambda_2 = e^{-i \alpha}$. Therefore
$1 + p \ep^2 = 1 + q \ep^2$, i.e., $q = p$. In the case $q = p$
the discriminant $\Delta$ of the quadratic equation \rf{quadr} is given by
\[
\Delta = - 4 \ep^2  + \ep^4 (1 - 4 p) \ .
\] 
There are two possiblities: if $p \geq 1/4$, then $\Delta < 0$ for any $\ep \neq 0$, 
and if $p < 1/4$, then $\Delta < 0$ for sufficiently small $\ep$, namely 
$\ep^2 < 4 (1 - 4 p)^{-1}$. 
In any case, 
these requirements are not very restrictive and we obtained $p$-family of good 
 discretizations of the harmonic oscilltor. 
If $\Lambda_1 = e^{i \alpha}$ and  $\Lambda_2 = e^{-i \alpha}$, then the solution of 
\rf{family} is given by
\be  \label{solfam}
x_n =  x_0 \cos (t_n \omega) + 
\frac{x_1 - x_0 \cos\alpha}{\sin\alpha}
\sin(t_n \omega) 
\ee
where $\omega = \alpha/\ep$, i.e., 
\be  \label{pomega}
  \omega = \frac{1}{\ep} \arctan \left( \ep \, \frac{ \sqrt{1 + \ep^2 \left( p - \frac{1}{4} \right)}}{1 + \left( p - \frac{1}{2} \right) \ep^2 } \right) \ .  
\ee
Note that the formula  \rf{solfam} is invariant with respect to the transformation $\alpha \rightarrow - \alpha$ which means that we can choose as $\Lambda_1$ any of the two roots
of \rf{quadr}. 

The equation \rf{dis-sym-os} is a special case of \rf{family} for $q = p = 0$. As we have seen in Section~\ref{three}, for small $\ep$ this discretization simulates the harmonic oscillator \rf{oscyl} much better 
than \rf{dis-Euler-os} or \rf{dis-forw-os}. However, for sufficiently large $\ep$ (namely, $\ep > 2$) 
the properties of this discretization  change
dramatically. Its generic solution 
grows exponentially without oscillations.

Expanding \rf{pomega} in the Maclaurin series with respect to $\ep$ we get
\be
\omega \approx 1 + \frac{1 - 12 p}{24} \ \ep^2 + 
\frac{3 - 40 p + 240 p^2}{640} \ \ep^4 + \ldots 
\ee
Therefore the best approximation of \rf{oscyl} from among the family \rf{family} is characterized by 
$p = 1/12$:
\be  \label{bestpq}
x_{n+1} - 2 \left( \frac{ 12 - 5 \ep^2}{12 + \ep^2} \right) x_n + x_{n-1} = 0 \ . 
\ee 
In this case $\omega \approx  1 + \ep^4/480 + \ldots$ is closest to 
the exact value $\omega = 1$.

The standard numerical methods 
give  similar results (in all cases presented in Appendix~\ref{app-numer} the discretization of the second derivative is the simplest one, the same as described in our Introduction). 
The corresponding discrete equations do not simulate \rf{oscyl} better than the discretizations presented in Section~\ref{three}.

\section{Damped harmonic oscillator and its discretization}.

\begin{figure}[p] 
\caption{\small Simplest discretizations of the weakly damped harmonic oscillator equation 
($\omega_0=1$, $\gamma=0.1$) for small $t$
and $\ep=0.3$.
Black points: Eq.~\rf{damped-forw}, white points: Eq.~\rf{damped-sym}, 
grey points: Eq.~\rf{damped-back}, thin line: exact continuous solution. }

 \label{damped-simple}
 \includegraphics[width=\textwidth]{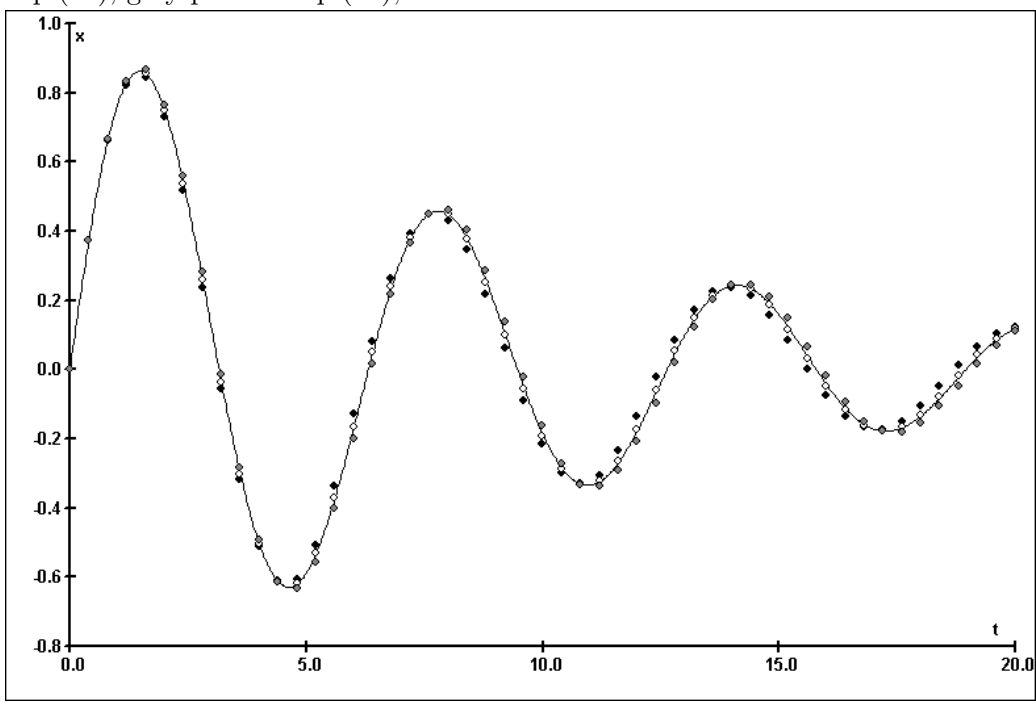} 

\end{figure}

Let us consider the damped harmonic oscillator equation \rf{damped-os}. Its general 
solution can be expressed by the roots $\lambda_1, \lambda_2$ of the characteristic equation
$\lambda^2 + 2 \gamma \lambda + \omega_0^2 = 0$ and the initial data $x (0)$, ${\dot x}(0)$:
\be  \label{cont-soll}
x(t) = \left( \frac{ {\dot x} (0) - \lambda_2 x(0)}{\lambda_1 - \lambda_2}
\right) e^{\lambda_1 t} + \left( \frac{ {\dot x} (0) - 
\lambda_1 x(0)}{\lambda_2 - \lambda_1}
\right) e^{\lambda_2 t} \ .
\ee
In the weakly damped case ($\omega_0 > \gamma > 0$) we have 
$\lambda_1 = - \gamma + i \omega$ and   $\lambda_2 = - \gamma - i \omega$, where
$\omega = \sqrt{\omega_0^2 - \gamma^2}$. Then
\be  \label{damped-sol}
  x (t) =  x(0) \ e^{- \gamma t} \cos \omega t + 
\omega^{-1} ( \dot x (0) + \gamma \, x (0) ) \ e^{-\gamma t} 
\sin \omega t \ .
\ee
To obtain some simple discretization of \rf{damped-os} we should replace the first 
derivative and the second derivative by discrete analogues. The results of Section~\ref{three} suggest that the best way to discretize the second derivative is the symmetric one, like 
in Eq.~\rf{dis-sym-os}. There are at least 3 possibilities for the discretization of the first 
derivative leading to the following simulations of the damped harmonic oscillator equation:
\be  \label{damped-forw}
\frac{ x_{n+1} - 2 x_{n} + x_{n-1} }{\ep^2} + 
2 \gamma \frac{x_n - x_{n-1}}{ \ep} + \omega_0^2 
 x_{n} = 0 \ . 
\ee
\be  \label{damped-sym}
\frac{ x_{n+1} - 2 x_{n} + x_{n-1} }{\ep^2} + 
2 \gamma \frac{x_{n+1} - x_{n-1}}{ 2 \ep} + \omega_0^2 
 x_{n} = 0 \ . 
\ee
\be  \label{damped-back}
\frac{ x_{n+1} - 2 x_{n} + x_{n-1} }{\ep^2} + 
2 \gamma \frac{x_{n+1} - x_n}{ \ep} + \omega_0^2 
 x_{n} = 0 \ . 
\ee
As one could expect, the best simulation is given by the most symmetric equation, i.e.,
Eq.~\rf{damped-sym}, see Fig.~\ref{damped-simple}.

\section{The exact discretization of the damped harmonic oscillator equation}

In order to find the exact discretization of \rf{damped-os} we consider the general linear discrete equation of second order
\be   \label{diseq-ab}
  x_{n+2} = 2 A x_{n+1} + B x_n \ .
\ee
The general solution of \rf{diseq-ab} 
has the following closed form (compare Appendix~\ref{app-lineq}):
\be  \label{dis-gensol}
   x_n = \frac{x_0 ( \Lambda_1 \Lambda_2^n - \Lambda_2 \Lambda_1^n ) +
x_1 ( \Lambda_1^n - \Lambda_2^n ) }{\Lambda_1 - \Lambda_2}
\ee
where $\Lambda_1, \Lambda_2$ are roots of the characteristic equation
$\Lambda^2 - 2 A \Lambda - B = 0$, i.e.,
\be \label{la12}
  \Lambda_1 = A + \sqrt{A^2 + B} \ , \qquad
  \Lambda_2 = A - \sqrt{A^2 + B} \ .
\ee
The formula \rf{dis-gensol} is valid for $\Lambda_1 \neq \Lambda_2$,
which is equivalent to $A^2 + B \neq 0$. If the eigenvalues 
coincide ($\Lambda_2 = \Lambda_1 $, $B = -A^2$) we have $\Lambda_1 = A$ and
\be  \label{coincide}
 x_n = (1-n) \Lambda_1^n x_0 + n \Lambda_1^{n-1} x_1 \ .
\ee

 {\it Is it possible to 
identify $x_n$ given by \rf{dis-gensol} with $x(t_n)$ where $x(t)$ is given by 
\rf{cont-soll}? }    \par
Yes! It is sufficient to express in an appropriate way $\lambda_1$ and $\lambda_2$ by
$\Lambda_1$ and $\Lambda_2$ and also 
the initial conditions $x(0), {\dot x}(0)$ by $x_0, x_1$. 
It is quite surprising that the above identification can be done 
for any $\ep$.

The crucial point consists in setting
\be
\Lambda_k^n  =  \exp (n \ln \Lambda_k)) = \exp (t_n \lambda_k) \ ,
\ee
where, as usual, $t_n : = n \ep$. It means that 
\be \label{lala}
\lambda_k : = \ep^{-1} \ln \Lambda_k \ ,
\ee
(note that for imaginary $\Lambda_k$, say $\Lambda_k = \rho_k e^{i \alpha_k}$, we have 
$\ln \Lambda_k = \ln \rho_k  +  i \alpha_k$).
Then \rf{dis-gensol} assumes the form
\be  \label{dis-soll}
  x_n = \left( \frac{x_1 - x_0 e^{\ep \lambda_2} }{e^{\ep \lambda_1} - 
e^{\ep \lambda_2} } \right) e^{\lambda_1 t_n} + 
\left( \frac{x_1 - x_0 e^{\ep \lambda_1} }{e^{\ep \lambda_2} - 
e^{\ep \lambda_1} } \right) e^{\lambda_2 t_n} \ .
\ee
Comparing \rf{cont-soll} with \rf{dis-soll} we get $x_n = x (t_n)$ provided that 
\be  \label{rellin}
x (0) = x_0 \ , \qquad
{\dot x} (0) = 
\frac{ (\lambda_1-\lambda_2) x_1 - (\lambda_1 e^{\ep \lambda_2} -
\lambda_2 e^{\ep \lambda_1} ) x_0 }{e^{\ep \lambda_1} - e^{\ep \lambda_2}}
\ . 
\ee
The degenerate case, $\Lambda_1 = \Lambda_2$ (which is equivalent to 
$\lambda_1 = \lambda_2$) can be considered analogically
(compare Appendix~\ref{app-lineq}). The formula \rf{coincide} is obtained from 
\rf{dis-gensol} in the limit $\Lambda_2 \rightarrow \Lambda_1$.
Therefore all formulas for the degenerate case can be 
derived simply by taking the limit $\lambda_2 \rightarrow \lambda_1$.

\begin{figure}[p] 
\caption{\small Good discretizations of the weakly damped harmonic oscillator equation 
($\omega_0=1$, $\gamma=0.1$) for small $t$
and $\ep=0.2$.
Black points: exact discretization \rf{dis-os-ex}, white points: Eq.~\rf{damped-sym}, 
grey points: Runge-Kutta scheme \rf{GLRK-harm}, thin line: exact continuous solution. }

 \label{damped-small}
 \includegraphics[width=\textwidth]{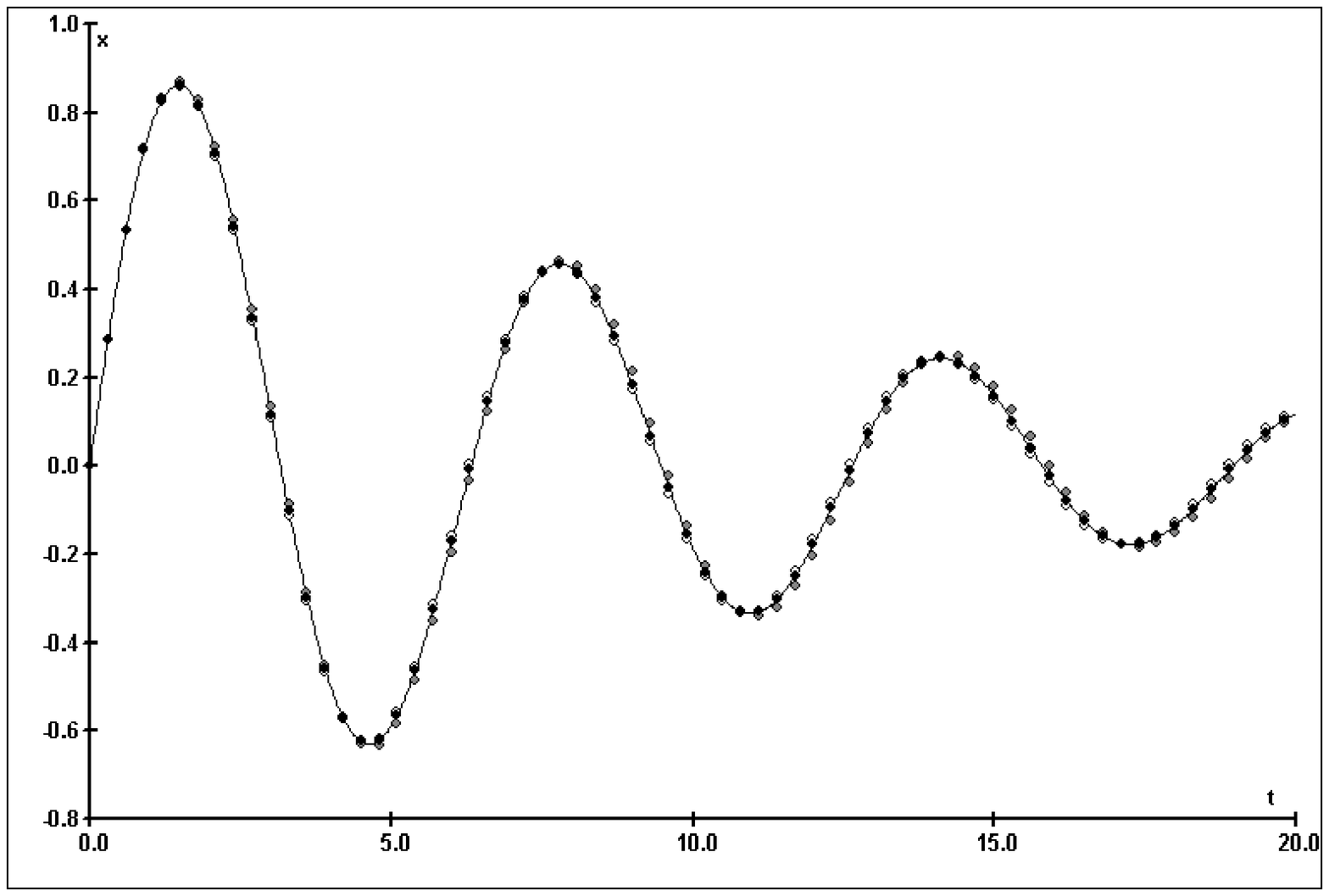} 

\end{figure}

\begin{figure}[p]  
\caption{\small Good discretizations of the weakly damped harmonic oscillator equation 
($\omega_0=1$, $\gamma=0.1$) for large $t$ and $\ep=0.2$.
Black points: exact discretization \rf{dis-os-ex}, white points: Eq.~\rf{damped-sym}, 
grey points: Runge-Kutta scheme \rf{GLRK-harm}, thin line: exact continuous solution. 
} 

\label{damped-large}
 \includegraphics[width=\textwidth]{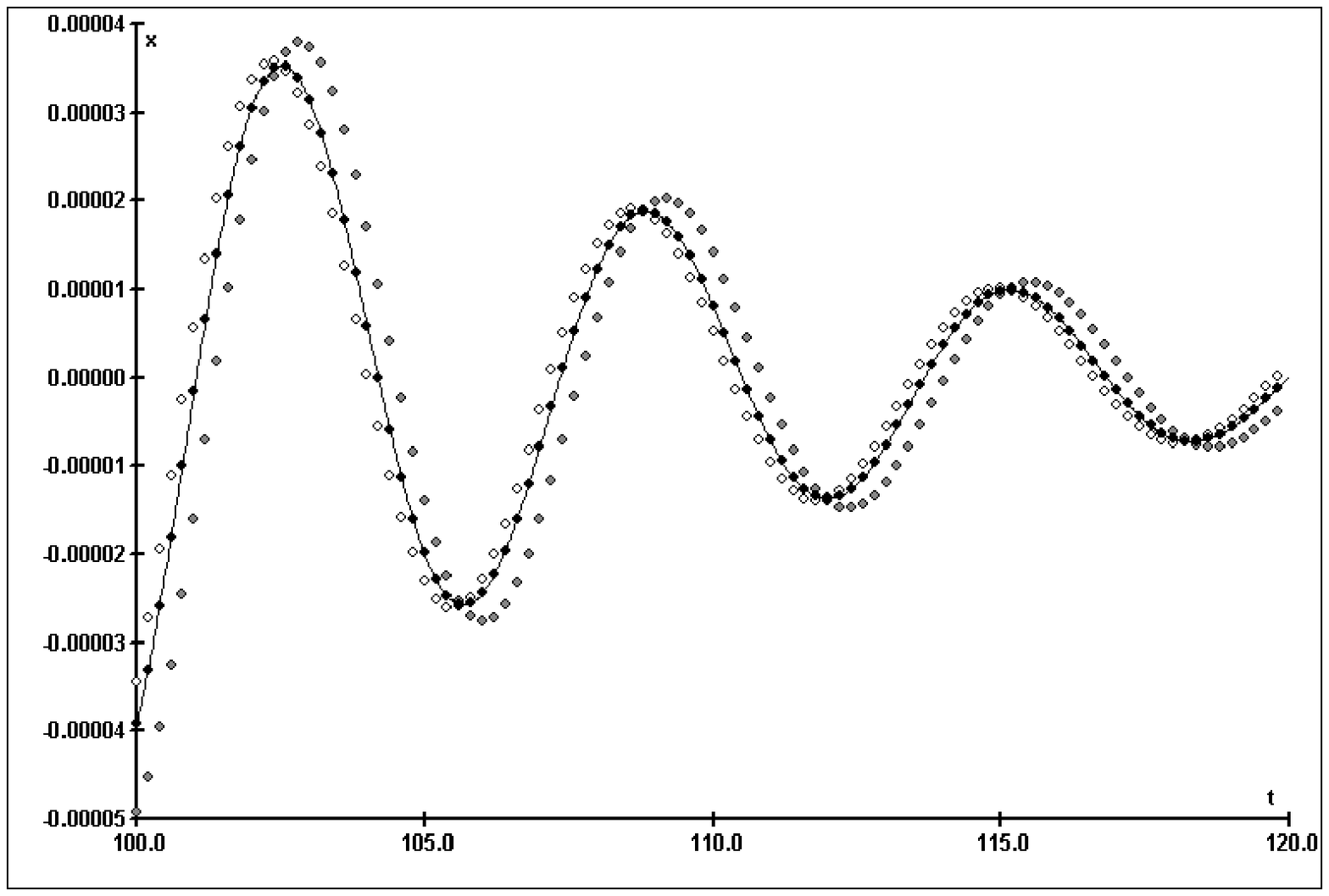} 

\end{figure}

Thus  we have a one-to-one
correspondence between second order differential equations with constant 
coefficients and second order discrete equations with constant coefficients.
This correspondence, referred to as the exact discretization, 
is induced by the relation \rf{lala} between the eigenvalues 
of the associated characteristic equations. 

The damped harmonic oscillator 
\rf{damped-os} corresponds to the discrete equation \rf{diseq-ab} 
such that
\be
2 A = e^{- \ep \gamma} \left( e^{\ep \sqrt{\gamma^2 - \omega_0^2} } + 
e^{ - \ep \sqrt{\gamma^2 - \omega_0^2} } \right) \ , \qquad 
B = - e^{ - 2  \ep \gamma} \ .
\ee

In the case of the weakly damped harmonic oscillator ($\omega_0 > \gamma >0$) the exact discretization is given by
\be
    A = e^{-\ep \gamma} \cos ( \ep \omega)  \ , \qquad
  B = - e^{-2 \ep \gamma} \ ,
\ee
where $\omega := \sqrt{\omega_0^2 - \gamma^2}$. In other words, 
the exact discretization of \rf{damped-os} reads
\be  \label{damped-exact}
x_{n+2} - 2 e^{-\ep \gamma } \cos(\omega\ep) x_{n+1} + 
e^{-2 \gamma \ep} x_n = 0 \ .
\ee
The initial data are related as follows (see \rf{rellin}):
\be  \ba{l}
x (0) = x_0 \ , \qquad \displaystyle {\dot x} (0) 
= \frac{x_1 \omega e^{\gamma \ep} -  (\gamma \sin (\omega \ep) +
\omega \cos (\omega \ep) ) x_0 }{\sin (\omega \ep)} \ ,  \\[4ex]
\displaystyle 
x_1 = \left( {\dot x} (0) \frac{\sin (\omega\ep)}{\omega} + 
\left( \gamma \frac{\sin (\omega\ep)}{\omega} + \cos (\omega\ep) \right)
x (0) \right) e^{- \ep \gamma} \ .
\ea \ee
Fig.~\ref{damped-small} and Fig.~\ref{damped-large} compare our exact discretization with 
two other good discretizations of the weakly damped harmonic oscillator.  
The exact discretization is really exact, i.e., the discrete points belong to the graph of 
the exact continuous solution (for any $\ep$ and any $n$). 
Similarly as in the 
undumped case, the fully symmetric simple discretization \rf{damped-sym} is better than 
the equation resulting from GLRK method. 

The exact discretization of the harmonic oscillator equation 
${\ddot x} + x = 0$ is a special case of \rf{damped-exact} and is given by
\be  \label{dis-os-ex}
  x_{n+2} - 2 (\cos\ep) x_{n+1} + x_n = 0 \ .
\ee
It is easy to verify that the formula \rf{dis-os-ex} can be rewritten as
\be
\frac{ x_{n+2} - 2 x_{n+1} + x_{n} }{ (2 \sin(\ep/2))^2} + x_{n+1} = 0 \ ,
\ee
which reminds the ``symmetric'' version of Euler's discretization scheme 
(see \rf{ddot} and \rf{dis-sym-os}) but $\ep$ appearing in  
 the discretization of the second derivative  is replaced by 
$2 \sin (\ep/2)$. For small $\ep$ we have $2 \sin (\ep/2)\approx \ep$. 

The comparison of the exact discretization \rf{dis-os-ex} 
with three other discrete equations 
simulating the harmonic oscillator is done in Fig.~\ref{good-smep}, 
and in Fig.~\ref{good-smt}.
We point out that the considered simulations are very good indeed 
(although $t$ in Fig~\ref{good-smep} is very large) but they cannot be better than 
the exact discretization.   The discretization \rf{bestpq} is also excellent. 
The coefficient by $- 2 x_n$ in Eq.~\rf{bestpq}  
\[
 \frac{12 - 5 \ep^2 }{12 + \ep^2} \approx 1 - \frac{1}{2!} \ep^2 + \frac{1}{4!} \ep^4 + \ldots
\]
approximates $\cos \ep$ up to $4$th order. Actually, for the choice of  parameteres 
made in Fig.~\ref{good-smep} and Fig.~\ref{good-smt}, the discretization \rf{bestpq} 
practically cannot be discerned from the exact one.

\section{Conclusions}

In this paper we have  shown that for linear ordinary differential equations 
of second order with constant coefficients 
 there exists a discretization which is ``exact'' and simulates properly 
all  features of the differential equation. 
The solutions of this  discrete equation exactly coincide 
with 
solutions of the corresponding differential equation evaluated at 
a discrete  lattice. Such exact discretization 
can be found for an arbitrary lattice spacing $\ep$.

Therefore we conclude that in this case differential and 
difference equations are in one-to-one correspondence: to any 
linear differential equation with constant coefficients 
there corresponds a difference equation which we call the 
exact discretization.

Analogical considerations can be made for linear ordinary differential 
equations (with constant coefficients) of any order (the details will be 
presented elsewhere).

We point out that to achieve our goal we had to assume an essential dependence 
of the discretization on the considered equation, in contrast to the standard numerical approach to ordinary differential equations where practically no assumptions are imposed on the considered system (i.e., universal methods, applicable for any equation, are considered) \cite{La}.

In the last years one can observe the development of the numerical methods 
which are applicable for some clases of equations (e.g., admitting Hamiltonian 
formulation) but  are much more powerful (especially when global or qualitative aspects are considered) \cite{St,IZ}. 

``Recent years have seen a shift in paradigm away from classical 
considerations which motivated the construction of numerical methods for 
ordinary differential equations. Traditionally the focus has been on the stability 
of difference schemes for dissipative systems on compact time intervals. 
Modern research is instead shifting in emphasis towards the preservation of 
invariants and the  reproduction of correct qualitative features'' 
\cite{Oe}.

In our paper we have a kind of extremal 
situation: the method is applicable for a very narrow class of equations but 
as a result we obtain the discretization which seems to be surprisingly good.

Similar situation occurs for the integrable (soliton) nonlinear systems. It is 
believed (and proved for a very large class of equations) that integrable equations 
admit integrable discretizations which preserve the unique features of these
equations (infinite number of conservation laws, solitons, transformations 
generating explicit solutions etc.) \cite{HA,BMS}.

The exact discretization considered in this paper  
is the best possible simulation of a differential equation. 
Linear ordinary differential equations always admit the unique exact discretization.
An open problem is to find such discretization for 
some other classes of differential equations.

\appendix

\section*{Appendix}

\section{Linear difference equations with constant coefficients}
\label{app-lineq}

We recall a  method of solving difference equations with 
constant coeficients. It consists in representing the equation 
in the form of a matrix equation of the first order.
The general linear discrete equation of the second order
\be   \label{second-dis}
  x_{n+2} = 2 A x_{n+1} + B x_n \ ,
\ee
can be rewritten in the matrix form as follows
\be \label{eee}
   y_{n+1} = M y_n \ , 
\ee 
where 
\be \label{MAB}
y_n = \m x_{n+1} \\ x_{n} \ema  \ , \qquad
 M = \mm 2A & B \\ 1 & 0 \ema \ .
\ee
The general solution of \rf{eee} has, obviously, the following form:
\be
   y_n = M^n y_0 
\ee
and the solution of a difference equation is reduced to 
the purely algebraic problem of computing powers of a given matrix.

The same procedure can be applied for any linear difference equation 
with constant coefficients. If the difference equation is of $m$-th order, 
then to obtain the equation \rf{eee} 
we define
\be
      y_n := (x_{n+m}, x_{m+m-1}, \ldots, x_{n+1}, x_n)^T
\ee
where the superscript $^T$ means the transposition. 

The power $M^n$ can be easily computed in the generic case in which 
the matrix $M$ can be diagonalized, i.e., represented in the form
\[
   M = N D N^{-1} \ ,
\]
where $D$ is a diagonal matrix. Then, obviously, $M^n = N D^n N^{-1}$.
The diagonalization is possible whenever the matrix $M$ has exactly 
$m$ linearly independent eigenvectors
(in particular, if the characteristic equation \rf{chareq} 
 has $m$ pairwise different roots). Then
the columns of the matrix $N$ are just the eigenvectors of $M$, 
and the diagonal diagonal coefficients of $D$ are eigenvalues of $M$.

The  characteristic equation   ($\det ( M - \lambda I) = 0 $) 
for $m=2$ (i.e., for \rf{second-dis}) has the form
\be  \label{chareq}
   \Lambda^2 - 2 A \lambda - B = 0 \ .
\ee
Its roots will be denoted by $\Lambda_1, \Lambda_2$ (see \rf{la12}). 
If $\Lambda_1 \neq \Lambda_2$, then the diagonalization procedure yields
\be
M  = 
N \mm \Lambda_1 & 0 \\ 0 & \Lambda_2 \ema  N^{-1}
\ee
where the columns of $N$ are the eigenvectors of M, i.e., 
\be
N = \mm \Lambda_1  & \Lambda_2 \\ 1 & 1 \ema \ .
\ee
Therefore
\be
 \m x_{n+1} \\ x_n \ema = 
N \mm \Lambda_1^n & 0 \\ 0 & \Lambda_2^n \ema  N^{-1}
     \m x_1 \\ x_0 \ema \ ,
\ee
and performing the multiplication we get \rf{dis-gensol}.

The case of multiple eigenvalues of $M$ is technically more 
complicated. In order to compute $M^n$ we can, for instance, 
transform $M$ to the Jordan canonical form (see, for instance, \cite{Lang}).
Here we suggest a method which is very efficient for $2\times 2$ matrices.
By the Cayley-Hamilton theorem (\cite{Lang}) any matrix satisfies its 
characteristic equation. In the case of \rf{second-dis} it means that
$M^2 = 2 A \lambda M + B$.
In the case of the double root ($B = - A^2$) 
one can easily prove by induction
\be
 M^n =   (1-n) A^n M  + n A^{n-1} \ .
\ee  
Substituting it to \rf{eee} we get immediately \rf{coincide}.

\section{Numerical methods for ordinary differential equations}
\label{app-numer}

In this short note we give basic informations about some 
numerical methods for ordinary differential equations
and we appply them to case of harmonic oscillator equation \rf{oscyl}.

A system of linear ordinary differential equations (of any 
order) can always be rewritten as a single  matrix equation 
of the first order:
\be
      \dot y = S y \ ,
\ee
where the unknown $y$ is a vector and $S$ is a given matrix (in general
$t$-dependent).  Numerical methods are almost always (see \cite{La})
constructed for a large class of ordinary differential  equations 
(including nonlinear ones):
\be
  \dot y = f (t, y) \ .
\ee
We denote by $y_n$ a numerical approximant to the exact 
solution $y(t_n)$.

\subsubsection*{Euler's method}

\be
 y_{k+1} = y_k + \ep f (t_k, y_k) 
\ee

In this case the discretization of $\ddot x + x = 0$ is given by \rf{dis-Euler-os}.

\subsubsection*{Modified Euler's methods}

\be
 y_{k+1} = y_k + \ep f (t_k + \frac{1}{2} \ep, y_k + \frac{1}{2} 
\ep f (t_k, y_k) ) 
\ee

\be
 y_{k+1} = y_k +  \frac{1}{2}  \ep ( f (t_k, y_k)  +  
 f (t_k + \ep, y_k + \ep f(t_k, y_k) ) ) 
\ee

Both methods lead to the following discretization of $\ddot x + x = 0$:
\be
 \frac{ x_{n+1} - 2 x_{n} + x_{n-1} }{\ep^2} + x_n + \frac{1}{4} \ep^2 x_{n-1} = 0 \ .
\ee
The roots of the characteristic equation are imaginary and
\be
 |\Lambda_1| = |\Lambda_2| = \sqrt{1 + \frac{\ep^4}{4} }  \ .
\ee

\subsubsection*{1-stage Gauss-Legendre-Runge-Kutta method}

\be
 y_{k+1} = y_k + \ep f (t_k + \frac{1}{2} \ep, \frac{y_k + y_{k+1}}{2} ) 
\ee
The application of this numerical integration scheme yields the following discretization of 
the damped harmonic oscillator equation:
\be
\frac{ x_{n+1} - 2 x_{n} + x_{n-1} }{\ep^2} + 
2 \gamma \frac{x_{n+1} - x_{n-1}}{ 2 \ep} + \omega_0^2 
\frac{x_{n+1} + 2 x_{n} + x_{n-1}}{4} = 0 \ . 
\ee
In the case $\gamma = 0$, $\omega_0 = 1$ (i.e., $\ddot x + x = 0$), 
we have  
\be
 \Lambda_1 = \frac{2 + i \ep}{2 - i \ep} \ , \qquad \Lambda_2 = \frac{2 - i \ep}{2 + i \ep} \ . 
\ee

\subsubsection*{Adams-Bashforth extrapolation formula}

\be
y_{n+1} = y_n + \ep \sum_{j=0}^k b_{kj} f(t_{n-j}, y_{n-j}) 
\ee
where $b_{kj}$ are specially chosen real numbers. In particular:
$b_{10} = 3/2$, $b_{11} =  - 1/2$, $b_{20} = 23/12$, $b_{21} = - 4/3$, $b_{22} = 5/12$.

In the case $k=1$ we obtain (for $\ddot x + x = 0$):
\be \label{Adams-os}
\frac{ x_{n+1} - 2 x_{n} + x_{n-1} }{\ep^2} + \frac{9 x_{n-1} - 6 x_{n-2} + x_{n-3}}{4} = 0
\ee
and the characteristic equation reads
\[ 
\Lambda^4 - 2 \Lambda^3 + \left( 1+ \frac{9}{4} \ep^2 \right) \Lambda^2 
- \frac{3}{2} \ep^2 \Lambda + \frac{1}{4} \ep^2 = 0 \ .
\]
This is an equation 
of the 4th order (with no real roots for $\ep \neq 0$).


\end{document}